\documentclass[journal,draftcls, onecolumn, 12pt]{IEEEtran}
\pdfoutput=1

\usepackage[usenames,dvipsnames]{xcolor}
\usepackage[pdftex]{graphicx}

\usepackage{cite}
\usepackage{amsmath,amssymb,amsfonts}
\usepackage{algorithmic}
\usepackage{graphicx}
\usepackage{textcomp}
\usepackage{xcolor}
\usepackage{algorithm}
\ifCLASSOPTIONcompsoc
\usepackage[caption=false,font=normalsize,labelfon
t=sf,textfont=sf]{subfig}
\else
\usepackage[caption=false,font=footnotesize]{subfig}
\fi
\graphicspath{{./figs/}}
\def\BibTeX{{\rm B\kern-.05em{\sc i\kern-.025em b}\kern-.08em
    T\kern-.1667em\lower.7ex\hbox{E}\kern-.125emX}}

\begin{document}

\title{Real-Time Prediction of Delay Distribution in \\ Service  Systems using Mixture Density Networks
}

\author{Majid Raeis, Ali Tizghadam and Alberto Leon-Garcia  \\[20pt] 
Department of ECE, University of Toronto, Canada. \\[-5pt]
Emails: m.raeis@mail.utoronto.ca, ali.tizghadam@utoronto.ca  and alberto.leongarcia@utoronto.ca. }

\setcounter{page}{1}
\maketitle
\thispagestyle{plain}
\pagestyle{plain}

\begin{abstract}
Motivated by interest in providing more efficient services in customer service systems, we use statistical learning methods and delay history information to predict the conditional distribution of the customers' waiting times in queueing systems. From the predicted distributions, descriptive statistics of the system such as the mean, variance and percentiles of the waiting times can be obtained, which can be used for delay announcements, SLA conformance and better system management. We model the conditional distributions by mixtures of Gaussians, parameters of which can be estimated using Mixture Density Networks. The evaluations show that exploiting more delay history information can result in much more accurate predictions under realistic time-varying arrival assumptions.
\end{abstract}


\section{introduction}
Services are by definition intangible products that are experienced by the customers. Services may be defined in different contexts such as telecommunication, transportation, healthcare, banking, etc. Because of the intangible nature of services, quality of service mainly depends on the customers' experience of the received service. One of the important measures of the quality of service is the delay experienced by the customers. Any service system has a limited service capacity and therefore is not capable of providing service to all customers at the same time, during high-demand periods. As a result, customers usually have to wait in queues in order to receive service.  
In other words, servers are analogous to shared resources among customers, which cannot be utilized by all the customers at the same time.
In addition to the service capacity limitations of the system, time-variability and randomness of the demand are other important factors which can lead to queue formation.

Waiting time prediction for the new customers can be beneficial from both the customers' and the service providers' points of view. Service providers can use delay predictions for better management of the system by adaptive matching of the service capacity to the demand. On the other hand, waiting time predictions can be used for delay announcements to the customers, which will result in lower uncertainty about the waiting times, and therefore, higher customer satisfaction~\cite{i2018}.

\subsection*{Previous Work}
Waiting time prediction in service systems can be classified into two categories: queueing-theoretic methods and data-based methods~\cite{i2018}. Let us first begin with the queueing-theoretic methods.

One of the earliest work on predicting customer's waiting time in a multi-server queue is~\cite{whitt99}. This paper investigates the possibility of improving delay predictions by exploiting information about the system state, and the elapsed service time of the customers in service, under non-exponential service time assumptions. Following up on~\cite{whitt99}, Ibrahim and Whitt have studied the performance of alternative queue-length-based and delay-history-based predictors. Three types of delay history information which have been used widely in these papers are the delay of the last customer to enter service (LES), the elapsed waiting time of the customer at the head of the line (HOL) and the delay of the last customer to complete service (LCS). The real-time performance of delay-history-based predictors such as LES and HOL predictors are studied for the standard $GI/M/c$ queueing model in~\cite{iw2009a}.
 In~\cite{iw2009b}, Ibrahim and Whitt extend their analysis to queueing models with \emph{abandonments}. Particularly, they study the overloaded $GI/GI/c+GI$ model, where $+GI$ denotes i.i.d abandonment times with general distributions. 
It is shown that the queue-length predictor performs poorly in this regime, while HOL remains an effective estimator. 
The performance of the delay-history based predictors in multi-server queueing systems with \emph{non-stationary} arrivals is studied in~\cite{iw2011a}. It is shown that the delay-history based predictors can have significant estimation bias, particularly if the system goes through alternating overload and underload periods. Moreover, a refined delay estimator based on HOL is introduced in~\cite{iw2011a} to cope with time-varying arrivals in the $M(t)/GI/c+GI$ model, where $M(t)$ represents nonhomogeneos Poisson arrivals. Finally, \emph{time-varying capacity} is taken into account in~\cite{iw2011b}, where four new predictors have been introduced for the $M(t)/M/c(t)+GI$ queueing model.

Limitations of the queueing-theoretic analysis have resulted in recent interest in data-based methods such as machine-learning algorithms and data-mining techniques. These methods have been used for waiting time prediction in different contexts and more realistic settings such as healthcare and transportation systems. For instance, machine learning techniques have been used in~\cite{demir} to predict flight delays by exploiting available information from sensors in the airport. Combining process mining and queueing-theoretic results, \cite{sender} introduces queue-mining techniques for predicting waiting times in service systems. Ang et al.~\cite{ang} propose a new predictor, called Q-Lasso, which combines the Lasso method from statistical learning and fluid models from queueing theory. The authors consider waiting time prediction in emergency departments and use datasets from four hospitals.

Both of these methods have their own advantages and shortcomings. One of the main disadvantages of the queueing-theoretic methods is that the analysis can easily become intractable by considering more realistic assumptions or including more information sources in the prediction process. Particularly, consider delay-history-based prediction, which is the focus of this paper. The existing queueing-theoretic methods based on delay history information are often limited to exponential service time assumptions and stationary arrival times. Moreover, the refined delay-history-based predictors that have been introduced for time-varying arrival settings, such as the ones proposed in~\cite{iw2011a}, require knowledge of the model parameters including arrival rate, service rate and the number of servers, which might not be available and need to be estimated as well. Another shortcoming of the existing queueing-theoretic methods is their limitation in exploiting all the available information, such as delay history of multiple recent customers, instead of only focusing on a single customer's delay history (e.g., LES or LCS). On the other hand, the prediction method and the feature selection process in data-based predictions are usually specialized for a particular application and do not provide much insights about the importance of the features. Furthermore, uncertainty of the estimations and the distribution of the waiting times are other important pieces which are often missing in the literature.
 All these reasons motivated us to use statistical learning methods to study queueing models under more realistic assumptions, such as time-varying arrivals and non-exponential service times. 
 
The remainder of this paper is organized as follows. In Section~\ref{sys_model}, we describe the queueing system model and formulate the problems that we are going to study in this paper. We discuss some theoretical results on the conditional distribution of the delay given HOL information in Section~\ref{norm}. In Section~\ref{mdn}, we briefly review mixture density networks and discuss how it can be used to predict the conditional distribution of the delay. The evaluation of the proposed predictors are presented in Section~\ref{eval}. Finally, Section~\ref{con} presents the conclusions and the future work.

\section{System Model and Problem Setting}\label{sys_model}
Consider a multi-server queueing system with infinite queue size and $c$ homogeneous servers with FCFS service discipline.  We do not assume a specific distribution for service times or inter-arrival times and therefore, the arrival and service processes can have non-stationary distributions. 
 Let $w_i$ denote the observed waiting time (before entering service) of the $i$'th last customer who entered service. Based on this definition, $w_1$ represents the LES delay, i.e., $w_{1} = w_{LES}$. Furthermore, the random waiting time of a new arrival conditional on observed delay history of the last $h$ customers who entered service, i.e. $\bold{w}_h = (w_1, w_2, \cdots, w_h)$, is represented by $W(\bold{w}_h)$. 

Our goal is to predict the distribution of a new arrival's waiting time, given that the customer has to wait and an observed delay history of $\bold{w}_h = (w_1, w_2, \cdots, w_h)$, i.e., we are aiming to predict the conditional distribution $P(W | \bold{w}_h )$. Furthermore, we are interested in predicting a single-value prediction for $W(\bold{w}_h)$, which will be denoted by $\widehat{W}(\bold{w}_h)$ in the rest of the paper. In the case where the delay predictor directly uses delay of the last customer to enter service as its prediction, i.e., $W(\bold{w}_h)  \equiv w_1 = w_{LES}$, we get the LES estimator. However, we are primarily concerned with the MMSE predictions, which can be obtained as $\widehat{W}(\bold{w}_h) \equiv E[W(\bold{w}_h)]$. In other words, $E[W(\bold{w}_h)]$ minimizes the mean squared error (MSE) of the predictor, which is defined as 
\begin{equation}
\text{MSE}(\widehat{W}(\bold{w}_h)) \equiv E\left[\left(W(\bold{w}_h) - \widehat{W}(\bold{w}_h)\right)^2 \right].
\end{equation}
It should be mentioned that it is difficult to determine MMSE predictor theoretically, even for the simple case of $h=1$, and therefore, we use statistical learning methods to estimate the conditional mean of $W(\bold{w}_h)$.

Our approach for  estimating the conditional distribution of the waiting time is to use mixture density networks (MDNs)~\cite{bishop}. In particular, the MDN uses a mixture of Gaussians to estimate the conditional distribution of the waiting time as follows
\begin{equation}
P(W|\bold{w}_h) = \sum_{k=1}^K \pi_k(\bold{w}_h) \mathcal{N}(W|\mu_k(\bold{w}_h), \sigma_k^2(\bold{w}_h)), \label{eq:mixture}
\end{equation}
where $\pi_k(\bold{w}_h) \in (0, 1)$, $\mu_k(\bold{w}_h)$ and $\sigma_k^2(\bold{w}_h)$ denote the the mixing coefficient, mean and variance of the $k^{th}$ kernel, respectively, given delay history information $\bold{w}_h$. We will discuss this method in more detail in Section~\ref{mdn}.

As we mentioned earlier, an important reason for estimating distribution of the delay is to obtain probabilistic bounds instead of just making predictions. More specifically, we can define a probabilistic lower-bound ($w_{lb}$) and upper-bound ($w_{ub}$) as follows:
\begin{align}
P(W(\bold{w}_h) > w_{ub}) &\leq \varepsilon_{ub}, \label{ub_bound}\\
P(W(\bold{w}_h) < w_{lb}) &\leq \varepsilon_{lb}, \label{lb_bound}
\end{align}
where $\varepsilon_{ub}$ and $\varepsilon_{lb}$ are the violation probabilities for the upper-bound and the lower-bound, respectively. Since probabilistic upper bound provides a pessimistic estimation of the waiting time, it can be a good candidate for delay announcements to the customers.   Finally, confidence interval is one of the other statistics that will be used in this paper to measure the amount of uncertainty for each prediction. Since the confidence intervals will be used along with the MMSE predictions, we define the confidence interval for random waiting time $W(\bold{w}_h)$ with confidence level $P_{cl}$, as an interval with endpoints $ (E[W(\bold{w}_h)]-x, E[W(\bold{w}_h)]+ x)$ such that:
\begin{equation}
P(E[W(\bold{w}_h)]-x < W(\bold{w}_h) < E[W(\bold{w}_h)]+ x) = P_{cl} \label{eq:conf_int}
\end{equation}
 
\section{Normal Approximation}\label{norm}
In this section, we discuss some theoretical results on the conditional distribution of a new arrival's waiting time given the HOL delay. Here, we focus on the HOL delay information since it results in the same performance as using the LES delay, while it makes the analysis easier~\cite{iw2009a}. In particular, we consider the distribution of a customer's waiting time conditional on the HOL delay in the $M_t/GI/c$ queueing system, where $M_t$ represents the nonhomogeneous Poisson arrival process. 

\begin{figure}[t!]
\centering
\includegraphics[scale=0.45]{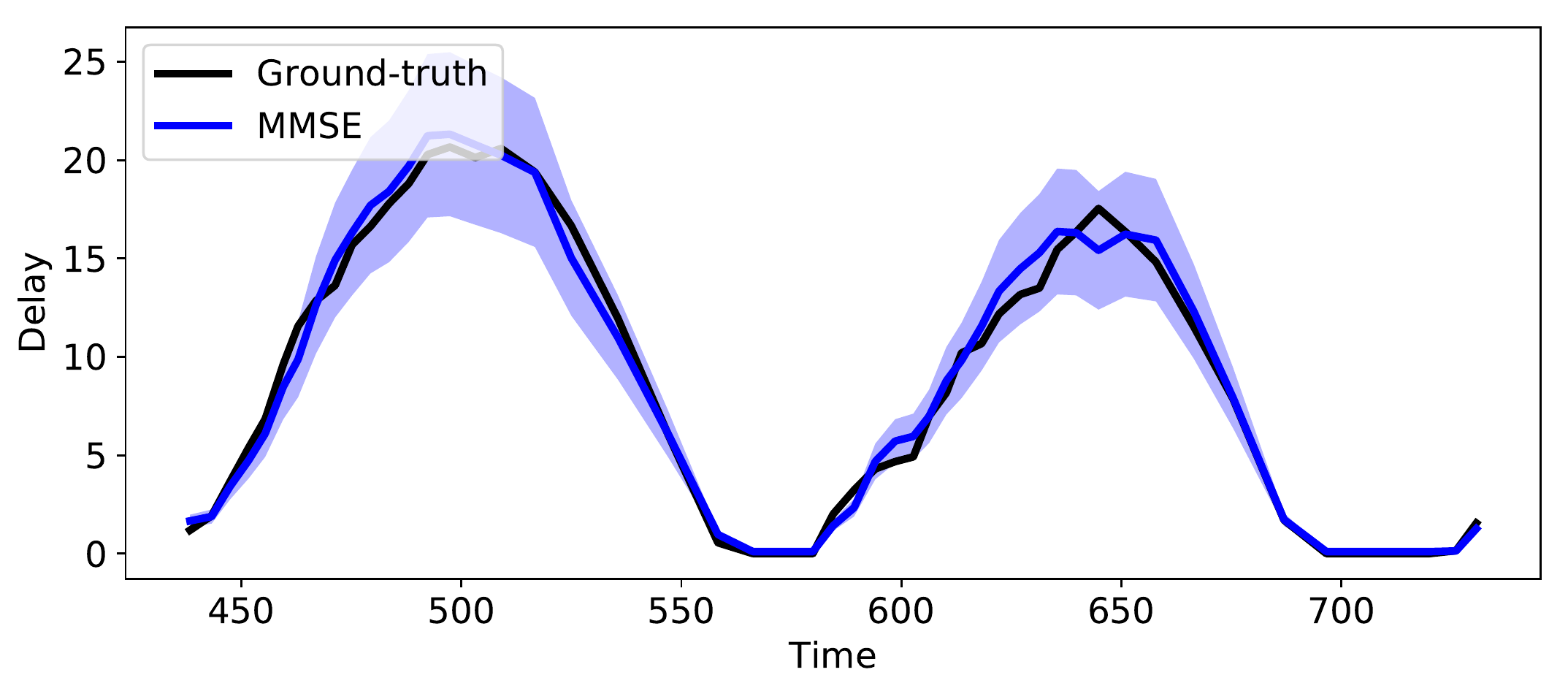}
    \caption{MMSE predictions along with 95\% confidence intervals, given the HOL information.}
    \label{fig:confInterval_theo}
\end{figure}

First, let us represent the delay of a new customer arriving at time $t$, given a HOL delay of $w_{HOL}$, by $W(t, w_{HOL})$, which can be written as
\begin{equation}
W(t, w_{HOL}) = \sum_{i=1}^{A(t)-A(t-w_{HOL})+2} S_i/c,
\end{equation}
where $A(t)$ denotes the number of arrivals in the interval $(0, t)$, and $S_i/c$ represents the time interval between successive service completions. Since the service times are i.i.d, the mean and the variance of $W(t, w_{HOL})$ can be obtained as follows
\begin{align}
E[W(t, w_{HOL})] &= \frac{E[A(t)-A(t-w_{HOL})]+2}{\mu c}, \label{eq:mean}\\
Var[W(t, w_{HOL})] &= \frac{E[A(t)-A(t-w_{HOL})]+Var[A(t)-A(t-w_{HOL})]+2}{(\mu c)^2}. 
\end{align}
For the proof please see~\cite{iw2009a}. Considering that the arrival process is modeled as a NHPP, the mean and the variance of the random variable $A(t)-A(t-w_{HOL})$ will be equal to $\int_{t-w_{HOL}}^{t} \lambda(\tau) d\tau$. Furthermore, since the conditional distribution of the delay given the past delay converges to the normal distribution when $cw_{HOL} \to \infty$ \cite{iw2009a}, the conditional mean and the variance often suffice to characterize the distribution of $W(t, w_{HOL})$. As an example, consider a system with nonhomogeneous Poisson arrivals and sinusoidal arrival rate as described in section~\ref{subsec:simExp}. Fig.~\ref{fig:confInterval_theo} shows a sample path of the ground-truth delay along with the MMSE predictions from Eq.~(\ref{eq:mean}). The filled region around the MMSE predictions show the region  where the ground-truth delays are expected to appear with a  confidence level of $95\%$, i.e., $P_{cl} = 0.95$. As can be observed, the uncertainty of the delay prediction increases as the delay becomes larger. 

Ibrahim and Whitt~\cite{iw2011a} have proposed a refined HOL-based delay estimator which uses the conditional mean calculated from Eq.~(\ref{eq:mean}) as its prediction. Although the refined HOL estimator performs better than the classical HOL estimator, especially in systems with time-varying arrivals, it requires additional information about the system parameters, as well as the arrival and service processes, which might not be available in practice and need to be estimated too. As a result, we are interested in estimators that only use the delay history information. However, instead of relying on the delay information of a single customer (HOL or LES delay), we use the delay-history of a larger number of past customers. Moreover, we use this information to estimate the conditional distribution of the delay using mixture density networks. 

\section{Mixture Density Networks (MDNs)}\label{mdn}

The mixture density network provides a general framework for approximating arbitrary conditional distributions. Using a mixture model as an approximation of the true conditional distribution, an MDN estimates the parameters of the mixture model using a fully-connected neural network. 
More specifically, the MDN approximates the conditional distribution of $\bold{y}$ given $\bold{x}$ by
\begin{equation}
P(\bold{y}|\bold{x}) = \sum_{k=1}^K \pi_k(\bold{x}) \mathcal{N}(\bold{y}|m_k(\bold{x}), \sigma_k^2(\bold{x})), \label{eq:mdn}
\end{equation}
where $\pi_k(\bold{x}) \in (0, 1)$ are the mixing coefficients and,  $m_k(\bold{x})$ and $\sigma_k^2(\bold{x})$ denote the mean and variance of the $k$'th kernel, $0 \leq k \leq K$, given  $\bold{x}$. Since the conditional distribution of the delay converges to the normal distribution under heavy-load assumptions, Gaussian kernels are reasonable candidates for the mixture model components.

The output layer of the neural network consists of three types of nodes which predict the parameters of the mixture model in Eq.~(\ref{eq:mdn}). The first type uses the soft-max activation function to predict the mixing coefficients such that $0 \leq \pi_k \leq 1$ and $\sum_k \pi_k =1$. The second group, which predict the variances of the kernels, use exponential activations to ensure non-negative values. The last group of the nodes use linear activations and compute the means of the kernels. Using a data set of $N_{sample}$ customers, where each sample contains $h$ features (delays of the last $h$ customers to enter service) and a label (delay of that particular customer), $\{(\bold{x}^j=\bold{w}_{h}^j, \bold{y}^j = W^j)|1 \leq j \leq N_{sample}\}$, the mixture density network learns the weights of the neural network by minimizing the error function, which is defined as the negative logarithm of the likelihood, i.e.,

\begin{equation}
E = - \sum_{j =1}^{N_{sample}} \ln \left\{ P(\bold{y}^j|\bold{x}^j)\right\}.
\end{equation}

It is worth mentioning that a standard feed-forward neural network with a single linear output unit trained by least squares, corresponds to maximum likelihood with a Gaussian distribution assumption~\cite{bishop}. As a result, the output of this simple neural network will approximate the conditional mean of the waiting time and therefore can be used to estimate the MMSE predictions. 

One of the main challenges in implementing a mixture density network is its instability issue.  In order to address this problem, various techniques and modifications have been proposed. An important problem associated with mixtures of Gaussians is the presence of singularities in the Likelihood function. In other words, in the case of having more than one Gaussian component, $K \geq 2$, the maximization of the Likelihood function is not a well-posed problem since the Likelihood can easily go to infinity whenever one of the Gaussian components lies on a specific data point and its variance goes to zero. For a more detailed discussion of the implementation issues related to mixture density networks refer to~\cite{mdn_thesis}.

\section{Evaluation and Results}\label{eval}

\begin{figure*}[t!]
\centering
\includegraphics[scale=0.4]{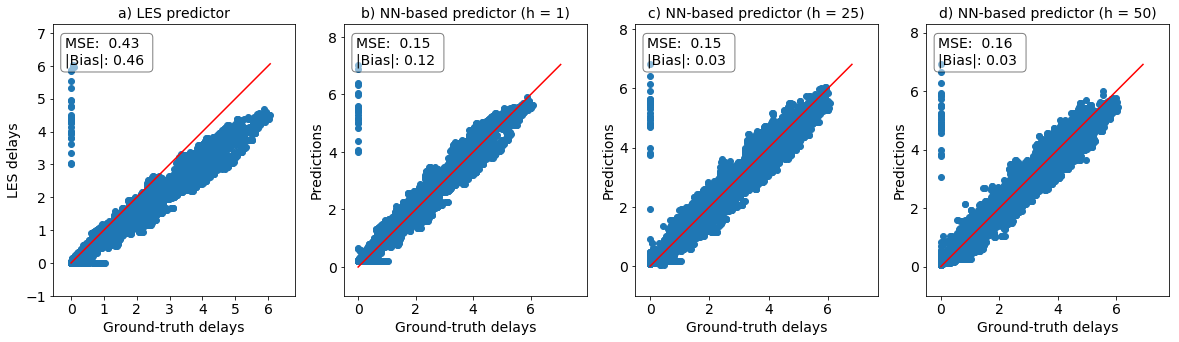}
    \caption{Scatter plots of the predicted delays and ground-truth delays}
    \label{fig:on-off_scatter}
\end{figure*}

\subsection{Performance Measures}
In order to evaluate the performance of the delay predictors, we consider two measures: \emph{absolute bias} and \emph{mean squared error} (MSE) for accuracy and precision, respectively. The absolute bias is defined as $\text{Bias}(\widehat{W}) = |E[W - \widehat{W}]|$ and will be approximated by 
\begin{equation}
\widehat{\text{Bias}} = \left|\frac{1}{N_{sample}} \sum_{i=1}^{N_{sample}} (d_i-p_i)\right|,
\end{equation}
where $d_i$ and $p_i$ are the ground-truth and predicted waiting times for the $i^{th}$ data point. 

The other measure is defined as $\text{MSE}(\widehat{W}) = E[(W-\widehat{W})^2]$ and will be approximated by the \emph{average squared error} as follows:

\begin{equation}
\text{ASE} = \frac{1}{N_{sample}} \sum_{i=1}^{N_{sample}} (d_i-p_i)^2.
\end{equation}

\subsection{Simulation Experiments}\label{subsec:simExp}
In the rest of this section, we present our results on delay prediction and distribution estimation in multi-server queueing systems. Let us begin with a short description of the arrival and service processes that are used in the following experiments. First, we consider a deterministic ON-OFF arrival process which can be used for simple approximation of a system with batch arrivals, such as transport terminals, in which arrival and departures occur in batches based on schedules. The ON-OFF arrivals have a cycle length of $4$ hours and a duty cycle equal to $75\%$. It should be noted that time is normalized to the mean service time in this paper. The second type of time-varying arrivals that are used in the experiments is the nonhomogeneous Poisson process, which is a good fit for arrivals in hospitals and call centers~\cite{nhpp}. We adopt the same model as in~\cite{iw2011a} with sinusoidal arrival rate to capture the daily cycles, i.e., we consider an arrival rate of $\lambda(t) = \bar{\lambda}(1+\alpha \sin(2 \pi t/T))$, where  $\bar{\lambda}$, $\alpha$ and $T$ represent the average arrival rate, relative amplitude and the cycle length of the arrival rate. We have evaluated systems with exponential, lognormal and H2 service times (hyperexponential with coefficient of variation equal to 2), however, the results are only presented for lognormal service times, due to space considerations.  Moreover, we use an MDN implementation which is based on~\cite{mdn_imp} and uses the Keras deep learning library. The training data set consists of around $27000$ samples and the test results are evaluated on $5000$ sample customers. As mentioned earlier, since we are studying delay-history-based predictors, we only consider delay of the last $h$ customers to enter service as the feature set. The simulation parameters are summarized in Table~\ref{ta:sim_table}.

\begin{table}[t!]
\centering
\caption{Parameters of the simulation.}
\begin{tabular}[t]{llr}
\hline
$\text{Notation}$ & Definition & value\\ 
\hline
$E[s]$ & Mean service time & 1 (10 mins)\\ 
$\rho$ & Traffic intensity & 0.95\\
$T$ & Cycle of NHPP arrivals & 144 (1 day)\\ 
$\alpha$ & Relative amplitude of $\lambda(t)$ & 0.5\\ 
$C$ & Number of servers & 20\\
\end{tabular}\label{ta:sim_table}
\end{table}

In the first experiment, we consider a multi-server queueing system with $20$ servers, deterministic ON-OFF arrival and lognormal service times. We use a simple feed-forward neural network to approximate the conditional mean of the delay given previous history, i.e., $E[W(\bold{w}_h)]$, and use it as our predictor. Fig.~\ref{fig:on-off_scatter} shows the scatter plots of the ground-truth and predicted delays for the LES predictor, and three other NN-based predictors with delay history lengths of $1, 25$ and $50$. As can be seen in Fig.~\ref{fig:on-off_scatter}.a, there exists a time lag between the LES delays and the ground-truth delays, particularly when the system is busy and the ground-truth values are large. On the other hand, we can observe that even a simple feed-forward neural network which only relies on the LES delay can achieve a much better performance and reduce the MSE by around $65\%$ compared to the LES predictor (see Fig.~\ref{fig:on-off_scatter}.b). Furthermore, the absolute bias has been reduced tremendously, which suggests that the systematic time lag does not exist anymore. It seems that increasing the delay history information does not have a noticeable impact on MSE in this experiment and might even result in overfitting and therefore a larger MSE (Fig.~\ref{fig:on-off_scatter}.d). From Fig.~\ref{fig:on-off_scatter} we can observe that the predictors can have large errors when the ground-truth delay is very small. One explanation might be that large past delays can either imply large delays for the next customers if the ON period continues, or suggest very small delays if the arrival process goes to the OFF state.
The vertical group of the outliers around zero ground-truth delay represent this phenomenon.

\begin{figure*}[t!]
\centering
\includegraphics[scale=0.4]{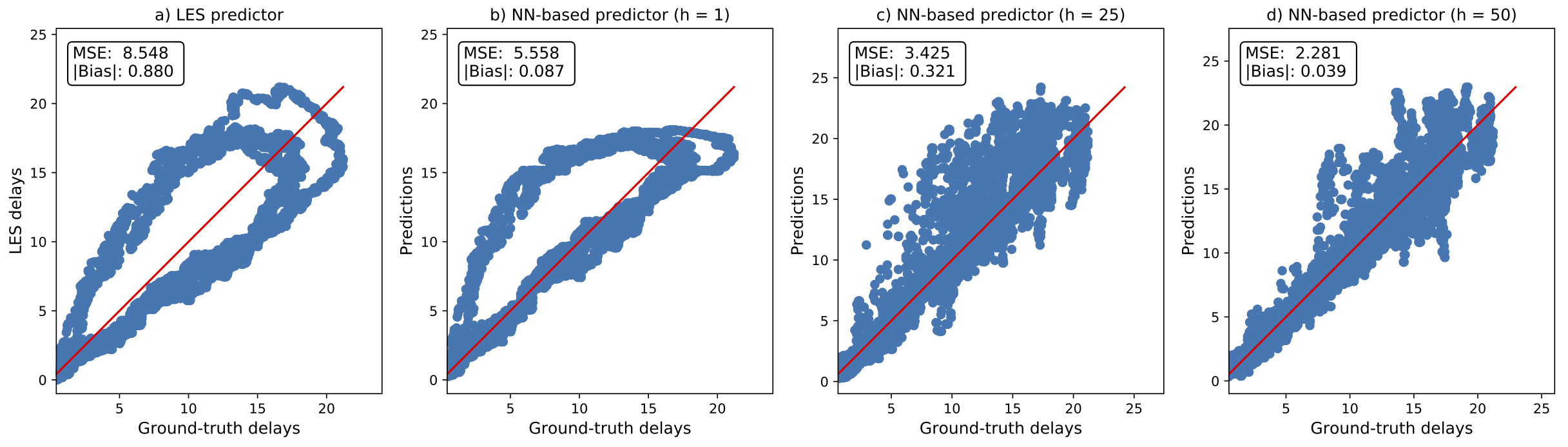}
    \caption{Scatter plots of the predicted delays and the ground-truth delays}
    \label{fig:nhpp_scatter}
\end{figure*}

\begin{figure}[t!]
\centering
\includegraphics[scale = 0.65]{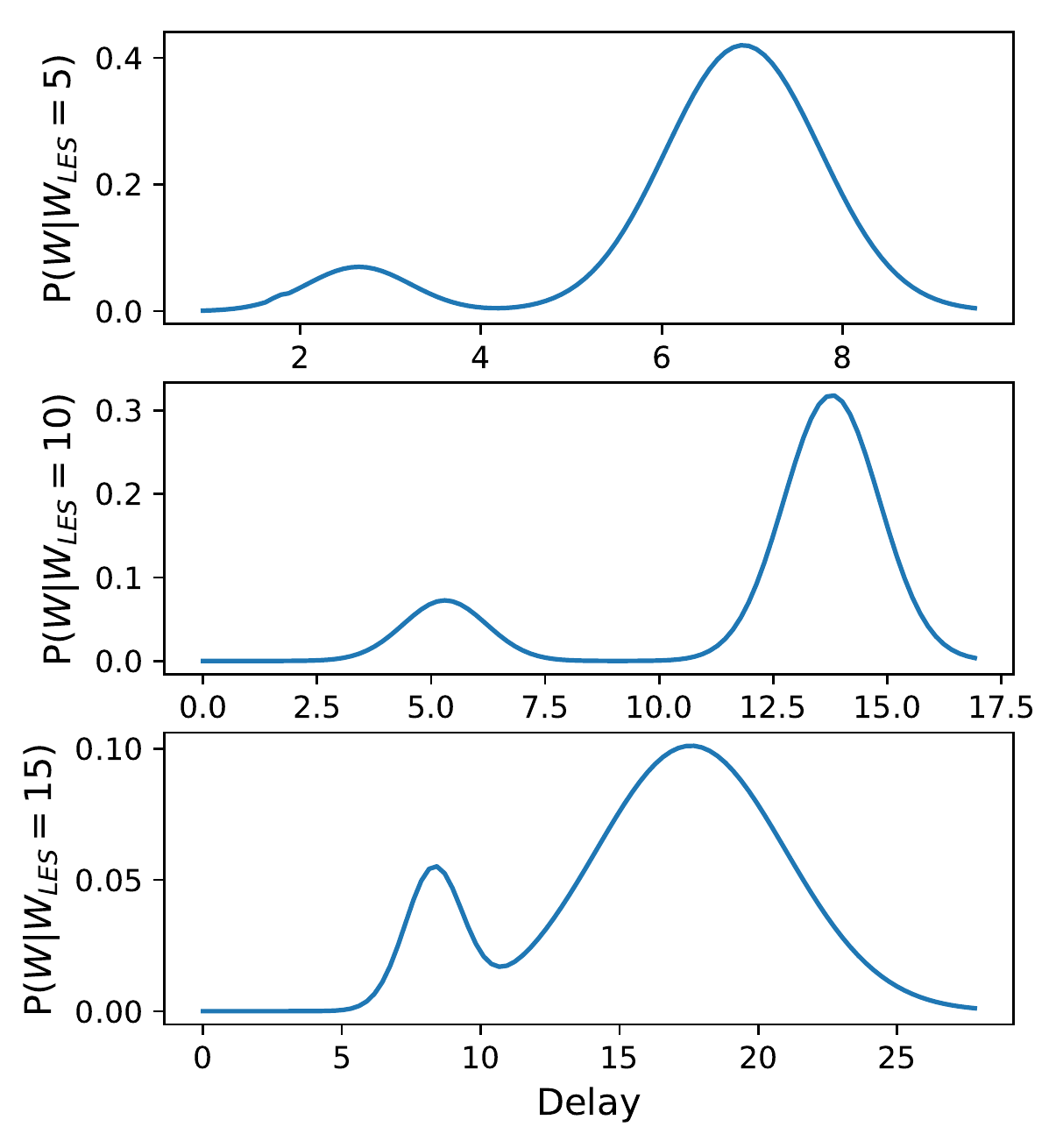}
    \caption{Estimated probability density function $P(W|w_1)$, for $w_1 = 5, 10$ and $15$. }
    \label{fig:pdfs}
\end{figure}

Now, consider the same queueing system but with NHPP arrivals. Fig.~\ref{fig:nhpp_scatter} shows the scatter plots for the same set of predictors. Similar to Fig.~\ref{fig:on-off_scatter}, there is a time lag between the LES delays and the ground-truth values. Moreover, Fig.~\ref{fig:nhpp_scatter}.a suggests that the conditional distribution of the delay given LES delay information should be multimodal. In other words, a particular LES delay can suggest both larger or smaller future delays, depending on whether the arrival rate and therefore system backlog, is in the increasing or decreasing phase. We observe that by increasing the length of the delay history to $h=50$, the NN-based predictor can better learn whether the system backlog is in the increasing or decreasing phase and hence, it is able to reduce the MSE by around $73\%$.

As we mentioned earlier, single value predictions of the waiting times might not be very informative and we are interested in more descriptive statistics of the system. For instance, it is more desirable to obtain stochastic upper bounds or lower bounds on a customer's delay similar to Eqs.~(\ref{ub_bound}) and (\ref{lb_bound}), rather than providing a single value prediction. In order to achieve this goal, we use MDNs as described in Section~\ref{mdn} to estimate the conditional distribution of the waiting time given delay history of the last $h$ customers to enter service. Fig.~\ref{fig:pdfs} shows the predicted conditional distribution of the waiting time in the previous experiment with NHPP arrivals, given LES delays equal to $5$, $10$ and $15$, i.e., $P(W|\bold{w}_1=5)$, $P(W|\bold{w}_1=10)$ and $P(W|\bold{w}_1=15)$, respectively. The two modes in the estimated distribution function, correspond to the two groups of points in the scatter plot shown in Fig.~\ref{fig:nhpp_scatter}.a, with the same LES delay value. Moreover,  Fig.~\ref{fig:pdfs} suggests that given a particular LES delay, it is more likely to have a larger delay for the new arrival. This can be explained by the fact that most of the customers with the same LES delay should have entered the system in the increasing phase of the arrival rate, while the queue length is growing, and therefore they are more likely to experience larger delays than the LES customer.

\begin{figure}[t!]
\centering
\includegraphics[scale=0.45]{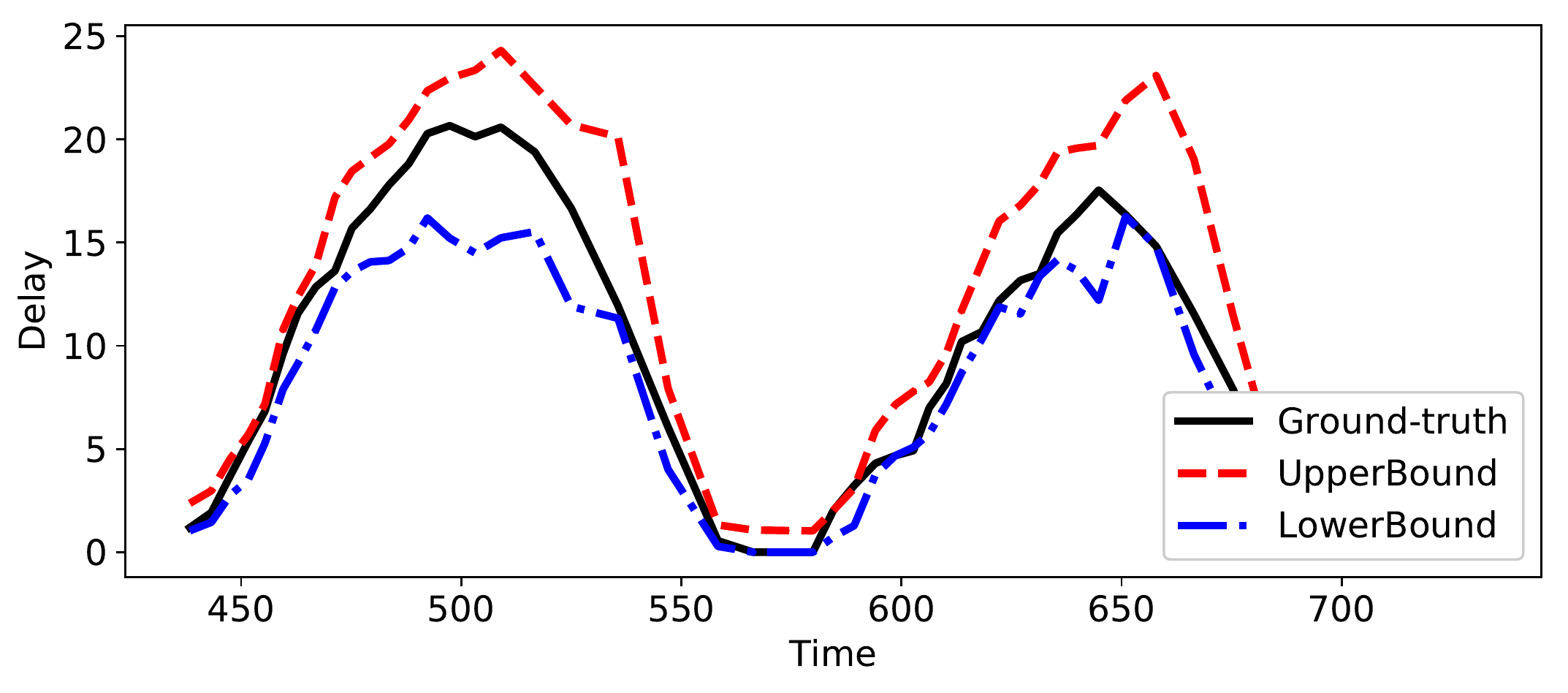}
    \caption{Probabilistic upper bound and lower bound on the waiting times with less than $5\%$ violation probabilities, given delay history of the last $h=50$ customers who entered service.}
    \label{fig:bounds}
\end{figure}

\begin{figure}[t!]
\centering
\includegraphics[scale=0.45]{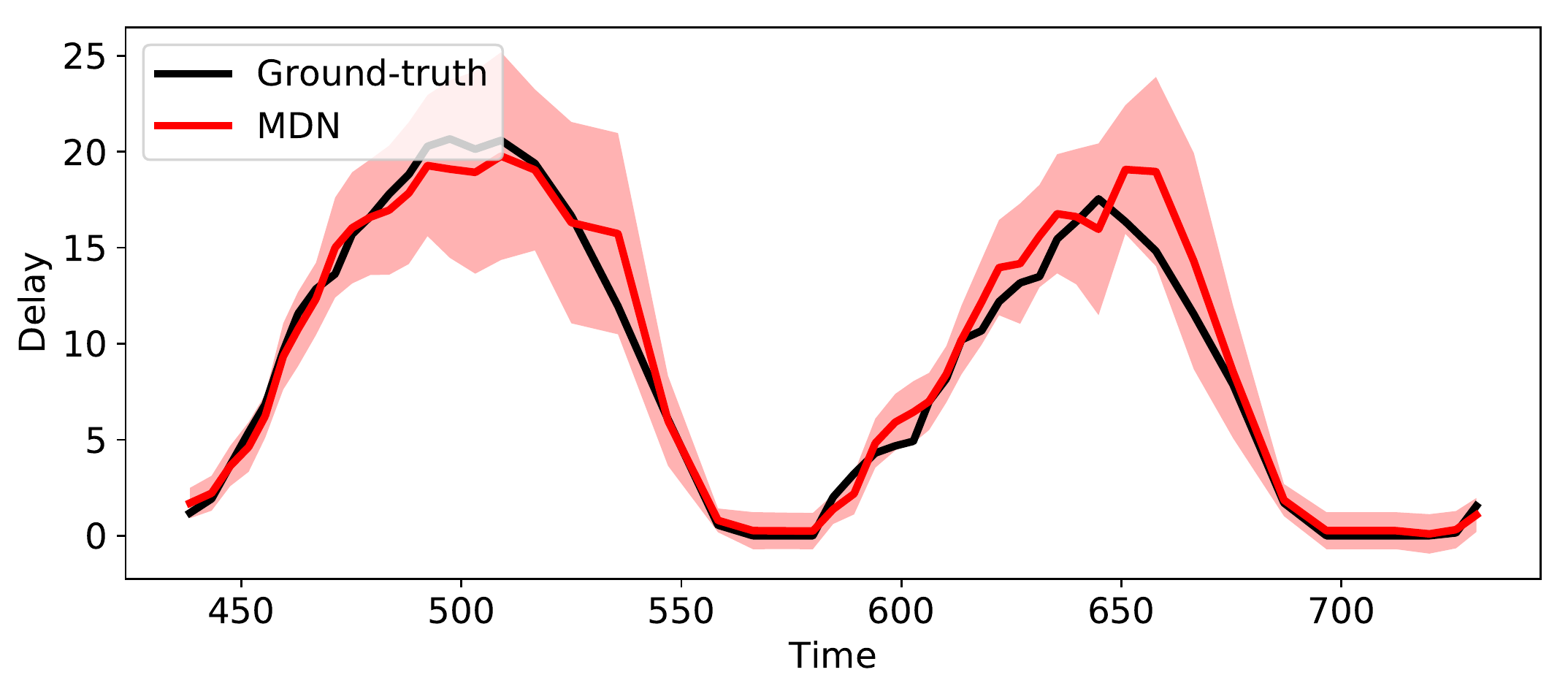}
    \caption{MMSE predictions along with 95\% confidence intervals (obtained from the MDN), given delay history length of $h=50$.}
    \label{fig:confInterval}
\end{figure}

Now, we use the estimated conditional distributions in a more effective way to obtain probabilistic bounds on the waiting time of a new arrival. Fig.~\ref{fig:bounds} shows a sample path of the waiting times in the previous experiment with NHPP arrivals for a period of two days. The stochastic upper bounds and lower bounds calculated from Eqs.~(\ref{ub_bound}) and (\ref{lb_bound}) are also shown in Fig.~\ref{fig:bounds}. The stochastic bounds are calculated to hold with a probability more than $0.95$, i.e, $\varepsilon_{ub}=\varepsilon_{lb}=0.05$. It should be mentioned that decreasing the violation probabilities can result in looser bounds. As mentioned earlier, the calculated upper bounds are good candidates for delay announcements. Closely related to these bounds, we can find the confidence intervals for each prediction using Eq.~(\ref{eq:conf_int}) and the predicted distributions. Fig.~\ref{fig:confInterval} shows the same sample path of the waiting times as in Fig.~\ref{fig:bounds}, along with the MMSE predictions and $95\%$ confidence intervals, obtained from the MDN. It can be observed that the confidence intervals can be pretty large for longer waiting times, which shows the importance of considering other statistics instead of focusing on single value predictions.

\section{Conclusions}\label{con}
In this paper, we attempted to show the potential of the statistical learning methods in providing insights on service systems under realistic assumptions, such as time-varying arrivals and non-exponential service times. In particular, we studied the problem of delay prediction and distribution estimation in multi-server queueing systems. We showed that even a very simple NN-based predictor that only uses the delay history information of the previous customers can outperform the traditional LES predictor (decreasing MSE by $73\%$), without requiring any other information about the system parameters. More importantly, MDNs enable us to estimate the conditional distribution of the waiting time, which can be used to obtain much more informative statistics, such as probabilistic bounds, compared to the common single-value predictions. 

Although the proposed NN-based methods are able to make good estimations for service systems under fixed model assumptions, we are interested in studying more realistic and complex cases, where model parameters such as the number of servers and service time distributions can change over time.
As discussed in the previous sections, these systems can appear in many different contexts, such as call centers and emergency departments, where theoretical analysis of the system under realistic assumptions gets intractable.

\vspace{12pt}

\end{document}